\documentclass[showpacs,%
 reprint,
 amsmath,amssymb,
 aps,
]{revtex4-1}

\usepackage{graphicx}
\usepackage{dcolumn}
\usepackage{bm}
\usepackage{hyperref}

\begin{document}

\preprint{APS/123-QED}

\title{Radiative Corrections in the (Varying Power)-Law Modified Gravity}

\author{Fay\c{c}al Hammad}
\email{fayhammad@gmail.com; fhammad@crc-lennox.qc.ca}
\affiliation{Physics Department \& STAR Research Cluster, Bishop's University,
\\\& Physics Department, Champlain College,\\
2600 College Street, Sherbrooke, Qu\'{e}bec J1M 1Z7, Canada.}


\begin{abstract}
Although the (varying power)-law modified gravity toy model has the attractive feature of unifying the early and late-time expansions of the Universe, thanks to the peculiar dependence of the scalar field's potential on the scalar curvature, the model still suffers from the fine-tuning problem when used to explain the actually observed Hubble parameter. Indeed, a more correct estimate of the mass of the scalar field needed to comply with actual observations gives an unnaturally small value. On the other hand, for a massless scalar field the potential would have no minimum and hence the field would always remain massless. What solves these issues are the radiative corrections that modify the field's effective potential. These corrections raise the field's effective mass rendering the model free from fine-tuning, immune against positive fifth-force tests, and better suited to tackle the dark matter sector.
\end{abstract}

\maketitle

\section{Introduction}\label{sec:1}
Modified gravity theories play a major role in cosmology \cite{CapozzielloFaraoni}. In the absence of any guiding principle one usually begins by writing down a simple toy model of modified gravity and uses it to tackle cosmological problems. One then tries to tailor one's model in such a way to make it capable of explaining, not only the early and the late-time expansions of the Universe (see e.g., \cite{Bamba, Nojiri1}), but also dark matter (see e.g., \cite{Capozziello1, Boehmer, Nojiri2, Stabile}).

Among the simplest and most attractive classes of modified gravity theories are the so-called power-law toy models. By adding extra terms with different powers of the scalar curvature inside the gravitational Lagrangian one makes gravity evolve with the Universe because these extra terms manifest themselves differently at different epochs in the evolution of the Universe \cite{Felice, Nojiri3}. However, when trying to fit many of the plausible models with observation one finds that the required powers of the Ricci scalar are not necessarily integers or rational numbers but might be real numbers that span a finite interval depending on the cosmological observation used in the fitting \cite{Carloni1, Capozziello1, Capozziello2}.

The (varying power)-law modified gravity toy model has been proposed in Ref.~\cite{Hammad} as a compromise between a power-law model and a scalar-tensor model \cite{Faraoni, Carloni2}. The power of the Ricci scalar in this model need not be fixed by hand each time one confronts the model with observation but rather left to be selected naturally by the surrounding environment. In Ref.~\cite{Hammad}, such a model was examined and found to be able to unify both the early and late-time expansions of the Universe as well as to eventually incorporate dark matter. The early expansion of the Universe might easily be incorporated within the model thanks to one of its two free mass parameters. However, when the model is used to represent the late-time expansion of the Universe, a fine-tuning issue arises because the second free mass parameter, which is supposed to represent the mass of the scalar field, is required to be unnaturally small in order to make the model able to reproduce the actually observed rate of cosmic expansion.

Now, although the scalar field is present in the model as the power of the Ricci scalar, which makes the model highly non-linear, quantum corrections due to the quantum fluctuations of the scalar field could easily be incorporated. Remarkably enough, it turns out that thanks to the peculiar form of the scalar field's potential, quantum fluctuations of the scalar field make the latter acquire an effective mass through its coupling with geometry such that no fine-tuning is required on the initial mass of the scalar field. Moreover, the mass thus acquired is so big that the model becomes immune against the fifth-force tests \cite{Bo}.

The outline of the paper is as follows. In Sec.~\ref{sec:2}, we give a more correct estimate of the scalar field's mass needed, before the quantum corrections are taken into account, to produce the currently observed Hubble parameter. We show that it should be, contrary to what has been suggested in Ref.~\cite{Hammad}, exceedingly small. We then argue why a completely massless field couldn't be used satisfactorily either. In Sec.~\ref{sec:3}, we show that despite the high non-linearity of the model, the application of the regular techniques in perturbation theory to take into account the quantum fluctuations of the scalar field is justified. We then compute the radiative corrections to the potential at one-loop order, find the needed counter terms, and finally write down the effective potential that would result from these radiative contributions to the classical potential. Using the obtained effective potential we deduce the expression of the effective mass of the scalar field. In Sec.~\ref{sec:4}, we show how the behavior of the effective potential is affected at late-times, making it possible to remedy the fine-tuning issue. We show that at low curvatures a massive as well as a massless scalar field both acquire a very big mass. The dynamical equation for the Hubble parameter is then given and the latter is found to be compatible with its actually observed value without imposing any kind of fine-tuning. We end this paper by a brief discussion section.

\section{The model and its fine-tuning issue}\label{sec:2}
Unless otherwise stated, the natural units $\hbar=c=1$ will be used through out the paper. Then, the gravitational action of the model studied in Ref.~\cite{Hammad} is
\begin{eqnarray}\label{1}
\frac{M^{2}_{\mathrm{P}}}{2}\int\mathrm{d}^{4}x\sqrt{-g}\left[R-\partial_{\mu}\phi\partial^{\mu}\phi
-m^{2}\phi^{2}-\mu^{2}\left(\frac{R}{R_{0}}\right)^{\phi}\right],\nonumber\\
\end{eqnarray}
where $M^{2}_{\mathrm{P}}$ stands for the Planck mass and, for definiteness, we have chosen the scalar field's kinetic energy parameter used in Ref.~\cite{Hammad} to be equal to one. The scalar field's mass is $m$ whereas $\mu$ is a mass parameter, taken in \cite{Hammad} to be of the order $\sim10^{1\sim10}\mathrm{GeV}$, but which we take here to be of the order of the GUT scale, namely $\sim10^{16}\mathrm{GeV}$ for reasons to be given below. The scalar curvature $R_{0}$ has been identified with the Planck curvature $M^{2}_{\mathrm{P}}\sim10^{38}(\mathrm{GeV})^{2}$.

A very important step before examining the consequences of having such a peculiar model is to isolate the scalar field's potential, which can be read off from (\ref{1}) as,
\begin{equation}\label{2}
V=\frac{1}{2}m^{2}\phi^{2}+\frac{1}{2}\mu^{2}\left(\frac{R}{R_{0}}\right)^{\phi}.
\end{equation}
Then a straightforward computation shows that the minimum of this potential occurs for a value $\phi_{0}$ of the scalar field given by \cite{Hammad}
\begin{equation}\label{3}
\left(\frac{R}{R_{0}}\right)^{\phi_{0}}=\frac{-2m^{2}\phi_{0}}{\mu^{2}\ln\frac{R}{R_{0}}}.
\end{equation}
With this identity at hand, we can go on to examine what Hubble flow would be produced, both in the early and late-time evolution of the Universe, whenever the field $\phi$ settles down at the bottom of its potential curve specified by this identity. For that purpose we shall take up the Friedmann equation given in Eq.~(13) of Ref.~\cite{Hammad} for a flat Friedmann-Lema\^{\i}tre-Robertson-Walker (FLRW) Universe:
\begin{multline}\label{4a}
\frac{\mu^{2}\phi}{R}\left(\frac{R}{R_{0}}\right)^{\phi}\dot{H}
+\left[1+\frac{\mu^{2}\phi}{R}\left(\frac{R}{R_{0}}\right)^{\phi}\right]H^{2}
\\=\frac{m^{2}\phi^{2}}{6}+\frac{\mu^{2}}{6}\left(\frac{R}{R_{0}}\right)^{\phi}+\frac{\dot{\phi}^{2}}{6}
+\mu^{2}H\frac{\mathrm{d}}{\mathrm{d}t}\left[\frac{\phi}{R}(R/R_{0})^{\phi}\right],
\end{multline}
where $H=\dot{a}/a$ is the Hubble parameter corresponding to the scale factor $a(t)$ of the flat FLRW metric. Given the high non-linearity of this dynamical equation, however, we will focus mainely on the orders of magnitude. In order therefore to make the analysis more tractable we shall simplify the above equation and write it in the following form,
\begin{multline}\label{4b}
\frac{\mu^{2}\phi}{R}\left(\frac{R}{R_{0}}\right)^{\phi}\dot{H}
+\left[\alpha+\frac{\mu^{2}\phi}{R}\left(\frac{R}{R_{0}}\right)^{\phi}\right]H^{2}
\\=\frac{m^{2}\phi^{2}}{6}+\frac{\beta\mu^{2}}{6}\left(\frac{R}{R_{0}}\right)^{\phi},
\end{multline}
where $\alpha$ and $\beta$ are dimensionless constants of order unity. To obtain the latter form, we have transposed the term $\dot{\phi}^{2}/6$ from the right-hand side of (\ref{4a}) to the left-hand side and noticed that $\dot{\phi}$ is of the order $H$, and hence, the quadratic term $\dot{\phi}^{2}/6$ is of the order $H^{2}/6$. Also, the term $\mu^{2}H\frac{\mathrm{d}}{\mathrm{d}t}[\frac{\phi}{R}(R/R_{0})^{\phi}]$ being of the same order as the term $\mu^{2}(R/R_{0})^{\phi}/6$, both present in the right-hand side of (\ref{4a}), we have approximated the former by the latter and multiplied by the factor of order  unity $\beta$. We shall now examine the consequences of equations (\ref{3}) and (\ref{4b}) successively for the early and late-time expansion of the Universe.

As for the early Universe, due to the prevailing high curvatures, the Ricci scalar $R$ might be taken to be of the order of the Planck curvature $R_{0}$ making the ratio $R/R_{0}$ very close to unity and hence the logarithm in (\ref{3}) very small. Furthermore, as argued in \cite{Hammad}, the scalar field starts out very close to the origin, $\phi\ll1$, making the term $(R/R_{0})^{\phi_{0}}$ in the left-hand side of (\ref{3}) very close to unity. This could only be consistent with the right-hand side however if the order of magnitude of the mass $m$ does not exceed the order of magnitude of $\mu$. Keeping only the dominant terms in (\ref{4b}) we would then be left with $H^{2}\sim\mu^{2}$, which implies that the Hubble flow squared $H^{2}_{\mathrm{Inf}}$ at the inflationary period would be of the order $\mu^{2}$ as already showed in \cite{Hammad}. Therefore the natural order of magnitude of $\mu$ could reach up to the GUT scale $\sim10^{16}\mathrm{GeV}$ according to what is required from the inflationary scales \cite{Zong, Liddle}.

Let us now use the two previous equations to examine the late-time expansion of the Universe, for which we shall take the Ricci scalar $R$ of the order of the actually observed Hubble parameter $H^{2}_{\mathrm{Obs}}\sim 10^{-66}(\mathrm{eV})^{2}$. In this case the ratio $R/R_{0}$ becomes of the order $\sim10^{-122}$. Therefore, for values of $\phi_{0}$ of the order $\sim10^{-105}$, as given in \cite{Hammad} in Planck unites, the required mass $m$ to satisfy (\ref{3}) should be absurdly big (of the order $\sim10^{72}\mathrm{eV}$ \cite{Hammad}) but moreover this would be in conflict with the late-time expansion. Indeed, keeping only the leading terms in (\ref{4b}) in this case would imply again that $H^{2}_{\mathrm{Obs}}\sim\mu^{2}$. In order to find the right estimate of $\phi_{0}$ at late times, and hence, the required mass $m$ let us substitute (\ref{3}) into (\ref{4b}). The latter then reads
\begin{multline}\label{5}
\frac{\mu^{2}\phi_{0}}{R}\left(\frac{R}{R_{0}}\right)^{\phi_{0}}\dot{H}
+\left[\alpha+\frac{\mu^{2}\phi_{0}}{R}\left(\frac{R}{R_{0}}\right)^{\phi_{0}}\right]H^{2}
\\=\frac{\mu^{2}}{6}\left(\frac{R}{R_{0}}\right)^{\phi_{0}}\left(\beta-\frac{\phi_{0}}{2}\ln\frac{R}{R_{0}}\right).
\end{multline}
Now it is clear that, given the orders of magnitude of $\mu^{2}$ and $R$ chosen above, what is needed is to have the term $(R/R_{0})^{\phi_{0}}$ tame the order of magnitude in the right-hand side of the above equation coming from the factor $\mu^{2}$. This could be achieved only if $\phi_{0}\sim1$. When substituting this back into (\ref{3}) we find in fact a very small but non-vanishing value for the scalar field's mass: $m\sim10^{-33}\mathrm{eV}$. This value is unnaturally small for a scalar field and hence the model is plagued with the usual fine-tuning problem.

Actually, a massless scalar field could very well be compatible with (\ref{4b}). In that case, the resulting Hubble parameter for values of $\phi_{0}$ very close to the origin would again be given by $H^{2}_{\mathrm{Inf}}\sim\mu^{2}$. Then, as $\phi$ departs from the origin, the term $\mu^{2}\phi(R/R)^{\phi}/R$ becomes gradually negligible and (\ref{4b}) would eventually reduce to
\begin{align}\label{6}
H^{2}\sim\mu^{2}\left(\frac{R}{R_{0}}\right)^{\phi_{0}}.
\end{align}
This would give the observed value $H^{2}_{\mathrm{Obs}}$ for the value $\phi_{0}\sim0.95$ of the scalar field. The issue in this case however lies in the fact that for a massless scalar field $\phi$, the potential (\ref{2}) would not have any minimum for finite values of $\phi$. Therefore, the scalar field would always remain massless and the model becomes vulnerable against the fifth-force problem, unless one imposes a non-coupling between the scalar field $\phi$ and matter, a property which is not desirable if one wishes to explain the creation of matter and radiation at the end of inflation.

A possible solution to this issue would be to modify the scalar field's potential in such a way that the minimum condition (\ref{3}) combined with (\ref{4b}) would allow us either (i) to have a reasonable and not fine-tuned mass $m$ for the field $\phi$ if the latter is chosen to be a massive scalar field, or (ii) to make the scalar field acquire a mass through it's coupling with gravity even if it was initially chosen to be massless. Such a modification, as we shall see in the next section, arises naturally by taking into account the quantum corrections.

\section{The effective potential from one-loop quantum corrections}\label{sec:3}
In this section we shall examine how radiative corrections modify the effective potential of the model. First, recall that quantum corrections to the potential inside a Lagrangian $\mathcal{L}$ are obtained from the following formal expansion around the classical value $\phi_{c}(x)$ of the scalar field given by the vacuum expectation value $\phi_{c}(x)=(\langle0|\phi(x)|0\rangle/\langle0|0\rangle)_{J}$ of the operator $\phi(x)$ in the presence of the source $J(x)$ (see e.g. \cite{Peskin}),
\begin{multline}\label{7}
\int(\mathcal{L}[\phi]+J\phi)=\int{(\mathcal{L}[\phi_{c}]+J\phi_{c})}+\int{\eta\left(\frac{\delta\mathcal{L}}{\delta\phi}+J\right)_{\phi_{c}}}
\\+\frac{1}{2}\int{\eta^{2}\frac{\delta^{2}\mathcal{L}}{\delta\phi^{2}}\Big|_{\phi_{c}}}
+\frac{1}{3!}\int{\eta^{3}\frac{\delta^{3}\mathcal{L}}{\delta\phi^{3}}\Big|_{\phi_{c}}}
+\frac{1}{4!}\int{\eta^{4}\frac{\delta^{4}\mathcal{L}}{\delta\phi^{4}}\Big|_{\phi_{c}}}+...,
\end{multline}
where $\eta(x)$ is the perturbation of $\phi(x)=\phi_{c}(x)+\eta(x)$ around the classical value $\phi_{c}(x)$ and all the functional derivatives of the Lagrangian with respect to the field $\phi$ are evaluated at the classical configuration $\phi_{c}$. Therefore, before we proceed with the usual method of finding the effective potential based on the effective action \cite{Peskin} we should first ascertain that such expansion would be convergent for the specific Lagrangian we have.

At first sight though it seems that the high non-linearity of the non-minimal coupling in the potential (\ref{2}) would render the expansion badly divergent and hence unjustified. When substituting the classical potential (\ref{2}) inside $\mathcal{L}$ in the above expansion, however, one finds simply the following formal expansion
\begin{align}\label{8}
\int(\mathcal{L}[\phi]&+J\phi)=\int{(\mathcal{L}[\phi_{c}]+J\phi_{c})}
+\int{\eta\left(\frac{\delta\mathcal{L}}{\delta\phi}+J\right)_{\phi_{c}}}\nonumber
\\&+\frac{1}{2}\int{\eta^{2}\left[-\partial^{2}+m^{2}
+\frac{\mu^{2}}{2}\left(\ln\frac{R}{R_{0}}\right)^{2}\left(\frac{R}{R_{0}}\right)^{\phi_{c}}\right]}\nonumber
\\&+\frac{\mu^{2}}{2}\frac{1}{3!}\int{\eta^{3}\left(\ln\frac{R}{R_{0}}\right)^{3}\left(\frac{R}{R_{0}}\right)^{\phi_{c}}}\nonumber
\\&+\frac{\mu^{2}}{2}\frac{1}{4!}\int{\eta^{4}\left(\ln\frac{R}{R_{0}}\right)^{4}\left(\frac{R}{R_{0}}\right)^{\phi_{c}}}+...
\end{align}
Then it is clear that, by redefining $\eta\rightarrow\eta'=\eta\ln(R/R_{0})$, the expression becomes a convergent series in the field $\eta'$. So the above expansion could be truncated as usual at the quadratic order in $\eta'$ making the use of the definition of the effective action, and hence of the radiative potential with contributions restricted to the one-loop diagrams, amply justified.

The vertex Feynman diagrams giving rise to such a non-minimal coupling of the scalar field $\phi$ with gravity are depicted in the figure below.
\begin{figure}[h]
\center{\includegraphics[angle=0, scale=0.45]{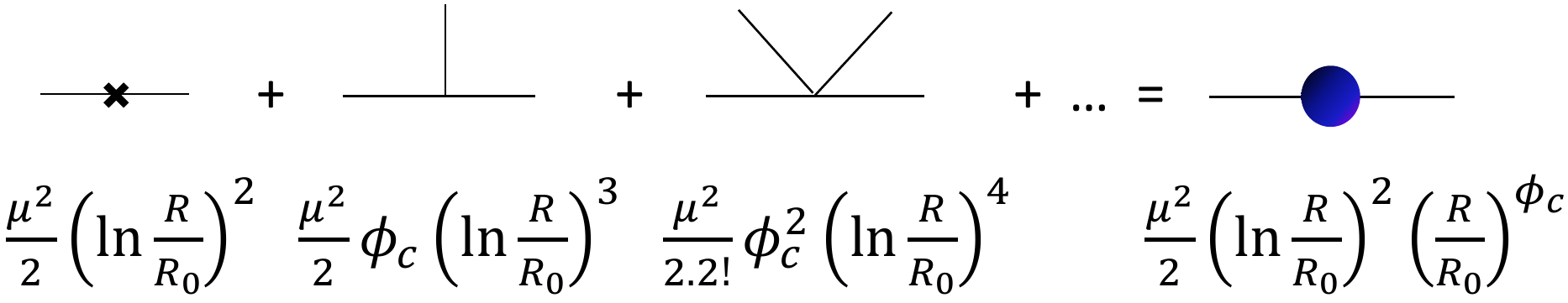}}
  \caption{\small{Vertex Feynman diagrams giving rise to the non-minimal coupling of the scalar field $\phi$ with gravity.}}\label{Vertex}
\end{figure}

Using these vertex diagrams, the one-loop radiative corrections to the effective potential, at which we restrain ourselves in this paper, are given by the following diagrammatic expansion:
\begin{figure}[h]
\center{\includegraphics[angle=0, scale=0.46]{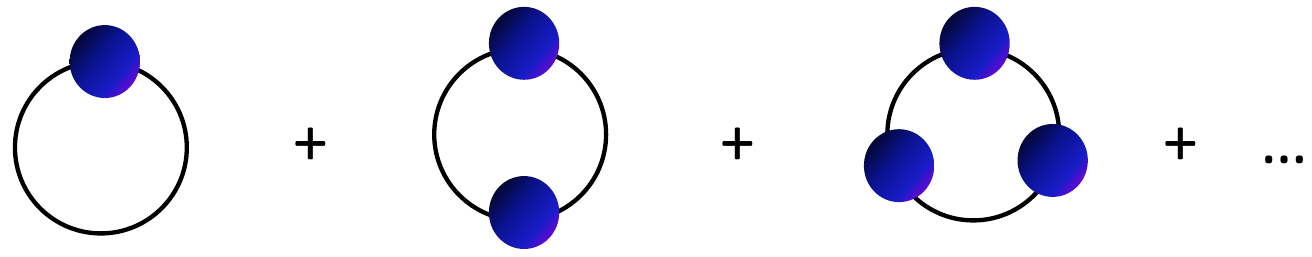}}
  \caption{\small{Diagrammatic expansion of the one-loop radiative contributions to the scalar field's potential.}}\label{Loop}
\end{figure}

Therefore, the radiative potential coming from this one-loop expansion would be given by the following integral (see e.g. \cite{Cheng})
\begin{equation}\label{9}
V=\frac{i}{M^{2}_{\mathrm{P}}}\int{\frac{\mathrm{d}^{4}k}{(2\pi)^{4}}}\sum^{\infty}_{n=1}
\frac{1}{2n}\left[\frac{\mu^{2}}{2}\left(\ln\frac{R}{R_{0}}\right)^{2}\frac{(R/R_{0})^{\phi_{c}}}{k^{2}-m^{2}+i\epsilon}\right]^{n}
\end{equation}
in which we have used the fact that the Feynman propagator is massive with mass $m$. After a Wick rotation, the sum may be transformed into a logarithm and the above integral becomes
\begin{equation}\label{10}
\frac{1}{16\pi^{2}M^{2}_{\mathrm{P}}}\int^{\Lambda}_{0}\mathrm{d}k_{E}k_{E}^{3}\ln\left[
1+\frac{\mu^{2}}{2}\left(\ln\frac{R}{R_{0}}\right)^{2}\frac{(R/R_{0})^{\phi_{c}}}{k_{E}^{2}+m^{2}}\right],
\end{equation}
where $\Lambda$ is the ultraviolet cut-off and $k_{E}$ is the Euclidean momentum. The evaluation of this last integral gives the radiative potential as,
\begin{align}\label{11}
&V=\frac{\mu^{2}}{128\pi^{2}M^{2}_{\mathrm{P}}}\left(\ln\frac{R}{R_{0}}\right)^{2}\left(\frac{R}{R_{0}}\right)^{\phi_{c}}\times\nonumber
\\&\left(\Lambda^{2}-\ln\frac{\Lambda^{2}}{M^{2}_{\mathrm{P}}}\left[2m^{2}+\frac{\mu^{2}}{2}\left(\ln\frac{R}{R_{0}}\right)^{2}\left(\frac{R}{R_{0}}\right)^{\phi_{c}}\right]\right)\nonumber
\\&+\frac{1}{64\pi^{2}M^{2}_{\mathrm{P}}}\left[m^{2}+\frac{\mu^{2}}{2}\left(\ln\frac{R}{R_{0}}\right)^{2}\left(\frac{R}{R_{0}}\right)^{\phi_{c}}\right]^{2}\times\nonumber
\\&\ln\left[\frac{m^{2}}{M^{2}_{\mathrm{P}}}+\frac{\mu^{2}}{2M^{2}_{\mathrm{P}}}\left(\ln\frac{R}{R_{0}}\right)^{2}\left(\frac{R}{R_{0}}\right)^{\phi_{c}}\right]-\frac{m^{4}}{64\pi^{2}M^{2}_{\mathrm{P}}}\ln \frac{m^{2}}{M^{2}_{\mathrm{P}}}.
\end{align}

Therefore, the effective potential when quantum corrections are included becomes
\begin{equation}\label{12}
V_{\mathrm{eff}}=\frac{m^{2}}{2}\phi^{2}+\frac{\mu^{2}}{2}\left(\frac{R}{R_{0}}\right)^{\phi_{c}}+V+\delta V,
\end{equation}
where $\delta V$ represents the contribution of the counter terms that should be added in order to cancel the divergences that rise in $V$ when the limit $\Lambda\rightarrow\infty$ is performed. Comparing (\ref{11}) and (\ref{12}), we deduce that the counter terms contribution should be of the following form,
\begin{equation}\label{13}
\delta V=A\left(\frac{R}{R_{0}}\right)^{\phi_{c}}+B\left(\frac{R}{R_{0}}\right)^{2\phi_{c}},
\end{equation}
where,
\begin{equation}\label{14}
A=\frac{-\mu^{2}}{128\pi^{2}M^{2}_{\mathrm{P}}}\left(\ln\frac{R}{R_{0}}\right)^{2}\left(\Lambda^{2}-2m^{2}\ln\frac{\Lambda^{2}}{M^{2}_{\mathrm{P}}}\right),
\end{equation}
\begin{equation}\label{15}
B=\frac{\mu^{4}}{256\pi^{2}M^{2}_{\mathrm{P}}}\left(\ln\frac{R}{R_{0}}\right)^{4}\ln\frac{\Lambda^{2}}{M^{2}_{\mathrm{P}}}.
\end{equation}
Note that although the form of the potential rising from the quantum corrections has a new form which we do not find in the original classical potential (\ref{2}), the number of counter terms required to cancel the divergences agrees with the number of free parameters the original potential had, namely, the two mass parameters $m$ and $\mu$.

Substituting these counter terms back inside the corrected potential (\ref{12}), the effective potential $V_{\mathrm{eff}}$ becomes finite and reads
\begin{align}\label{16}
V_{\mathrm{eff}}&=\frac{m^{2}}{2}\phi^{2}_{c}+\frac{\mu^{2}}{2}\left(\frac{R}{R_{0}}\right)^{\phi_{c}}-\frac{m^{4}}{64\pi^{2}M^{2}_{\mathrm{P}}}\ln \frac{m^{2}}{M^{2}_{\mathrm{P}}}\nonumber
\\&+\frac{1}{64\pi^{2}M^{2}_{\mathrm{P}}}\left[m^{2}+\frac{\mu^{2}}{2}\left(\ln\frac{R}{R_{0}}\right)^{2}\left(\frac{R}{R_{0}}\right)^{\phi_{c}}\right]^{2}\times
\nonumber
\\&\ln\left[\frac{m^{2}}{M^{2}_{\mathrm{P}}}+\frac{\mu^{2}}{2M^{2}_{\mathrm{P}}}\left(\ln\frac{R}{R_{0}}\right)^{2}\left(\frac{R}{R_{0}}\right)^{\phi_{c}}\right].
\end{align}
We can see straightaway from (16) that the extra terms brought by the radiative modifications to the classical potential could be as high as the original terms we started with thanks to this highly non-linear coupling of the scalar field with curvature. This has been made possible by making the second mass parameter $\mu$ contribute substantially and in varying strengths depending on the different gravitational environments. Before examining in detail in the next section the consequences of such dependence on the environment for the early and late-time evolution of the Universe, we shall first write down here the condition that gives the minimum of this effective potential and determine the expression of the resulting effective mass.

The minimum of the effective potential (\ref{16}) occurs at $\left(\partial V_{\mathrm{eff}}/\partial\phi\right)_{\phi_{0}}=0$. Differentiating expression (\ref{16}) once with respect to $\phi$ yields the following constraint,
\begin{align}\label{17}
&m^{2}\phi_{0}+\frac{\mu^{2}}{2}\ln\frac{R}{R_{0}}\left(\frac{R}{R_{0}}\right)^{\phi_{0}}\nonumber
\\&+\frac{\mu^{2}}{64\pi^{2}M^{2}_{\mathrm{P}}}\left(\ln\frac{R}{R_{0}}\right)^{3}\left(\frac{R}{R_{0}}\right)^{\phi_{0}}\times\nonumber
\\&\left[m^{2}+\frac{\mu^{2}}{2}
\left(\ln\frac{R}{R_{0}}\right)^{2}\left(\frac{R}{R_{0}}\right)^{\phi_{0}}\right]\times\nonumber
\\&\left(\frac{1}{2}+\ln\left[\frac{m^{2}}{M^{2}_{\mathrm{P}}}+\frac{\mu^{2}}{2M^{2}_{\mathrm{P}}}\left(\ln\frac{R}{R_{0}}\right)^{2}\left(\frac{R}{R_{0}}\right)^{\phi_{c}}\right]\right)=0.
\end{align}
The effective mass of the scalar field can be found by writing $m^{2}_{\mathrm{eff}}=\left(\partial^{2}V_{\mathrm{eff}}/\partial\phi^{2}\right)_{\phi_{0}}$. Differentiating twice the effective potential (\ref{16}) and then using the constraint (\ref{17}), gives
\begin{align}\label{18}
m^{2}_{\mathrm{eff}}&=m^{2}\left(1-\phi_{0}\ln\frac{R}{R_{0}}\right)\nonumber
\\&+\frac{\mu^{4}}{128\pi^{2}M^{2}_{\mathrm{P}}}\left(\ln\frac{R}{R_{0}}\right)^{6}\left(\frac{R}{R_{0}}\right)^{2\phi_{0}}\times\nonumber
\\&\left(\frac{3}{2}+\ln\left[\frac{m^{2}}{M^{2}_{\mathrm{P}}}
+\frac{\mu^{2}}{2M^{2}_{\mathrm{P}}}\left(\ln\frac{R}{R_{0}}\right)^{2}\left(\frac{R}{R_{0}}\right)^{\phi_{0}}\right]\right).
\end{align}

\section{Consequences of the effective potential}\label{sec:4}
Now that we have obtained all the necessary ingredients, we can look again at the consequences this effective potential brings to the dynamics of the Universe. To begin with, we shall first examine in which way the minimum condition (\ref{3}), found for a classical filed, would be modified here. Using (\ref{17}), we learn that the analogue of (\ref{3}) is given by
\begin{align}\label{19}
\left(\frac{R}{R_{0}}\right)^{\phi_{0}}=\frac{-2m^{2}\phi_{0}}{\xi\mu^{2}\ln\frac{R}{R_{0}}},
\end{align}
where the quantity $\xi$ is given by the following identity
\begin{multline}\label{20}
\xi=1+\frac{(\ln r)^{2}}{32\pi^{2}M^{2}_{\mathrm{P}}}\left(m^{2}+\frac{\mu^{2}(\ln r)^{2}}{2}x\right)\times
\\\left(\frac{1}{2}+\ln\left[\frac{m^{2}}{M^{2}_{\mathrm{P}}}+\frac{\mu^{2}(\ln r)^{2}}{2M^{2}_{\mathrm{P}}}x\right]\right),
\end{multline}
in which we set $r=R/R_{0}$ and defined $x=(R/R_{0})^{\phi_{0}}$.

First, for very high curvatures $r\sim1$, as might be the case during the early Universe, we have $\ln r\rightarrow0$ and therefore $\xi\sim1$, making thereby (\ref{19}) reduce to its classical counterpart (\ref{3}) whereas the effective potential (\ref{16}) recovers at this very high curvature its classical form (\ref{2}).

For very low curvatures $r\ll1$, however, thanks to the presence of this new factor $\xi$, the denominator in the right-hand side of (\ref{19}) could become big enough that the mass in the numerator would not have to be as small as what we found using the condition (\ref{3}). Therefore this new factor $\xi$ is exactly what would allow us to avoid the fine-tuning problem we found in the classical regime where $\xi$ reduces to unity. Moreover, thanks to this additional factor the scalar field could consistently be massless in (\ref{19}) provided only that $\xi=0$. Let us then first discuss the consequences of this change for the case of a massless scalar field and then examine the massive scalar field case.

\subsection{A Massless Scalar Field}\label{subsec:A}
In the case $m=0$, the minimum condition (\ref{19}) yields $\xi=0$ which, in turn, imposes the following constraint
\begin{align}\label{21}
\frac{1}{2}+\ln\left[\frac{x\mu^{2}(\ln r)^{2}}{2M^{2}_{\mathrm{P}}}\right]=-\frac{64\pi^{2}M^{2}_{\mathrm{P}}}{\mu^{2}(\ln r)^{4}x}.
\end{align}
Substituting this in the effective mass expression (\ref{18}), gives
\begin{align}\label{22}
m^{2}_{\mathrm{eff}}&=\frac{x^{2}\mu^{4}(\ln r)^{6}}{128\pi^{2}M^{2}_{\mathrm{P}}}
\left[1-\frac{64\pi^{2}M^{2}_{\mathrm{P}}}{\mu^{2}(\ln r)^{4}x}\right].
\end{align}
Therefore, although being massless initially, the scalar field acquires a huge mass coming from the mass parameter $\mu$ through geometry. Indeed, for low curvatures, by assuming $R\sim H^{2}_{\mathrm{Obs}}$ we have $r\sim10^{-122}$. Substituting this into (\ref{21}) and solving numerically for $x$ gives us actually two solutions, $x_{1}\sim0.014$ as well as $x_{2}\sim15$. The first solution is achieved for $\phi_{0}\sim0.01$ whereas the second solution is achieved for $\phi_{0}\sim-0.01$. Only the second solution gives a positive effective mass when using (\ref{22}) though. The first one would give a tachyon which means that the vacuum becomes unstable whenever $\phi$ departs from the origin towards the positive values. However, substituting the second solution into (\ref{22}) we find the remarkable fact that the mass acquired by the initially massless field when the latter departs from the origin towards the left of the vertical axis is of the order of $\sim10^{22}\mathrm{GeV}$.

Having a different effective potential, the dynamical equation for the Hubble parameter around the minimum $\phi_{0}$ of the effective potential will also be modified. In order to see this, first note that in terms of the effective potential, equation (\ref{4b}) actually reads,
\begin{equation}\label{23}
2\frac{\partial V_{\mathrm{eff}}}{\partial R}\Big|_{\phi_{0}}\dot{H}+\left(\alpha+2\frac{\partial V_{\mathrm{eff}}}{\partial R}\Big|_{\phi_{0}}\right)H^{2}\sim\frac{V_{\mathrm{eff}}}{3}\Big|_{\phi_{0}}.
\end{equation}
Next, substituting the above constraint (\ref{21}) into the effective potential $V_{\mathrm{eff}}$ given by (\ref{16}) as well as into its derivative $\partial V_{\mathrm{eff}}/\partial R$, the dynamical equation (\ref{23}) reads more explicitly as follows,
\begin{equation}\label{24}
-\frac{2x\mu^{2}}{R\ln r}\dot{H}+\left(\alpha-\frac{2x\mu^{2}}{R\ln r}\right)H^{2}\sim\frac{x\mu^{2}}{12}\left[1-\frac{x\mu^{2}(\ln r)^{4}}{128\pi^{2}M^{2}_{\mathrm{P}}}\right].
\end{equation}

Keeping only the leading terms in the left-hand side, this equation acquires for $x_{1}\sim0.014$ and $x_{2}\sim15$ the following two forms, $\dot{H}+H^{2}\sim\mp R\ln r$, respectively; the upper-sign corresponding to the first solution. Using the fact that in the flat FLRW geometry $R=6\dot{H}+12H^{2}$, these two possibilities are equivalent respectively to $H^{2}\sim\pm R\ln r$. Thus, we see that only the second solution corresponding to the minus sign, which is also the one that gives a stable vacuum as we saw, is acceptable. This, in turn, would give the observed Hubble parameter $H^{2}_{\mathrm{Obs}}\sim10^{-66}(\mathrm{eV})^{2}$ provided that $R\sim10^{-68}(\mathrm{eV})^{2}$. Note that although the solutions $x_{1}$ and $x_{2}$ obtained above for equation (\ref{21}) were found using the assumption $R\sim H^{2}_{\mathrm{Obs}}$, injecting the above slightly different estimate for $R$ again inside (\ref{21}) does not alter significantly the solutions $x_{1}$ and $x_{2}$ previously obtained because $R$ intervenes in (\ref{21}) only inside a logarithm. Therefore the order of magnitude for $H^{2}_{\mathrm{Obs}}$ that will be obtained from (\ref{24}) with these new solutions will be the same.

\subsection{A Massive Scalar Field}\label{subsec:B}
In the case $m\neq0$, the full minimum condition (\ref{19}), with $\xi$ given by (\ref{20}), also admits the previous two solutions $x_{1}\sim0.014$ and $x_{2}\sim15$. This being the case for all values of $m$ up to the GUT scale. This is due to the fact that the presence of $\mu^{2}$ multiplied by $\ln r$ overshadows the effect of $m^{2}$ inside $\xi$. When substituting the full minimum condition (\ref{19}) into expression (\ref{18}) of the effective mass, the latter becomes,
\begin{multline}\label{25}
m^{2}_{\mathrm{eff}}=m^{2}(1-\phi_{0}\ln r)+\frac{x^{2}\mu^{4}(\ln r)^{6}}{128\pi^{2}M^{2}_{\mathrm{P}}}\times
\\\left[1-\frac{64\pi^{2}M^{2}_{\mathrm{P}}}{x\mu^{2}(\ln r)^{3}}\cdot\frac{2m^{2}\phi_{0}+x\mu^{2}(\ln r)}{2m^{2}+x\mu^{2}(\ln r)^{2}}\right].
\end{multline}

First, we see that for the solution $x_{1}\sim0.014$ the vacuum will be unstable because the effective mass squared comes out negative again, whereas for the second solution $x_{2}\sim15$ the vacuum will be stable and the effective mass will be of the order $\sim10^{22}\mathrm{GeV}$, as found previously for the massless case. These results being true for any initial mass $m$ below the GUT scale. Inspecting the content of the square brackets in the above expression, however, we find that this time there is the possibility of having a stable vacuum for both positive and negative values of $\phi$, though at the price of starting with an initial mass $m$ that exceeds the GUT scale by one order of magnitude at least.

Finally, computing the effective potential $V_{\mathrm{eff}}$ as well as the derivative $\partial V_{\mathrm{eff}}/\partial R$ using the minimum condition (\ref{19}) we find, respectively,
\begin{align}\label{26}
V_{\mathrm{eff}}\Big|_{\phi_{0}}&=\frac{m^{2}\phi_{0}^{2}}{2}+\frac{x\mu^{2}}{4}
-\frac{m^{4}}{64\pi^{2}M^{2}_{\mathrm{P}}}\ln\frac{m^{2}}{M^{2}_{\mathrm{P}}}\nonumber
\\&-\frac{m^{2}}{x\mu^{2}(\ln r)^{3}}\left[m^{2}\phi_{0}+\frac{x\mu^{2}\ln r(1+\phi_{0}\ln r)}{2}\right]\nonumber
\\&-\frac{1}{128\pi^{2}M^{2}_{\mathrm{P}}}\left[m^{2}+\frac{x\mu^{2}(\ln r)^{2}}{2}\right]^{2},
\end{align}
and
\begin{align}\label{27}
\frac{\partial V_{\mathrm{eff}}}{\partial R}\Big|_{\phi_{0}}=-\frac{x\mu^{2}}{R\ln r}-\frac{(2+\phi_{0}\ln r)m^{2}\phi_{0}}{R(\ln r)^{2}}.
\end{align}
When inserting these new expressions inside (\ref{23}) we obtain again, after keeping only the leading terms, the same approximate dynamical equation (\ref{24}) for all values of the initial mass $m$ below the GUT scale. Then, the analysis and the conclusions drawn previously for a massless scalar field are also valid here. For an initially massive scalar field whose mass exceeds the GUT scale by at least one order of magnitude, for which the vacuum becomes stable on both side of the vertical axis, the terms containing $m^{2}$ dominate in both (\ref{26}) and (\ref{27}). Therefore, the latter expressions reduce to $V_{\mathrm{eff}}\big|_{\phi_{0}}\sim m^{2}\phi_{0}^{2}/2$ and $\partial V_{\mathrm{eff}}/\partial R\big|_{\phi_{0}}=-m^{2}\phi^{2}/(R\ln r)$, respectively. Inserting these inside (\ref{23}) we learn that the result $H^{2}\sim-R\ln r$ will be valid for both positive and negative values of $\phi$.

\section{Discussion}
The presence of the scalar field inside the power of the Ricci scalar in the (varying power)-law modified gravity model provides the latter with the main attractive feature of usual power-law models, namely, the unification of the early and late-time expansions of the Universe. Unfortunately though, the model also suffers from the usual fine-tuning problem familiar to anyone relying on scalar fields to explain the order of magnitude of the cosmological constant. The peculiar form of the scalar field's potential in this model, however, gives rise to substantial differences in the physics when the scalar field is treated quantum mechanically. The major modification brought by implementing the radiative corrections is the fact that not only the mass of the scalar field need not be finely adjusted to produce the effects of a tiny cosmological constant, but the mass of the field itself vary substantially, depending on the curvature of the surrounding environment. For low curvatures, the mass acquired by the scalar field is so big that the model could not in principle fail any fifth-force test. These conclusions being true for whatever initial mass the scalar field happens to have, nothing \textit{a priori} prevents us from identifying this scalar field with the fundamental Higgs field.

Another new feature found here, though, is the fact that, in contrast to the case one finds when the field is treated classically, the vacuum is not symmetric with respect to the vertical axis anymore. An asymmetry between the positive and negative values of the scalar field arises whenever the radiative corrections are taken into account. This asymmetry makes the vacuum unstable in the positive direction and stable in the negative direction. This asymmetry arises for all values of the initial mass of the scalar field that are below the GUT scale. For values exceeding the GUT scale by one order of magnitude and beyond, though, the vacuum becomes stable for both directions. Nevertheless, just by choosing minus the absolute value of the scalar field for the power of the Ricci scalar we could make the vacuum stable again for both directions.

In our investigation of the consequences of combining vacuum fluctuations of the scalar field with its peculiar non-minimal coupling with gravity we have completely ignored in this paper the matter sector and the eventual coupling of the matter fields with the scalar field. An important question arises then, which is: What difference would a coupling of the scalar field with matter bring to our results and conclusions. Although a rigorous analysis of this question lies beyond the scope of the present paper, hints for a qualitative answer could be found by recalling that whenever radiative corrections are taken into account, one obtains effective parameters depending on the precise structure of the interactions present in the model \cite{Peskin, Cheng}. Since in our model only the scalar field couples to gravity, the effects of vacuum fluctuations of the former due to its non-minimal coupling with the latter will appear simply as additive contributions to the usual calculations one performs in the absence of gravity. Indeed, taking into consideration a coupling of $\phi$ with matter fields $\psi$ within our model amounts to adding an interaction term $I(\phi,\psi)$ and the corresponding counter term $\delta I(\phi,\psi)$ in the sum on the right-hand side of expression (\ref{12}), giving the contributions of radiative corrections to the effective potential in the absence of gravity. This, in turn, would simply introduce in the right-hand side of (\ref{16}) an effective interaction potential $I_{\mathrm{eff}}(\phi,\psi)$. The consequences of having such additional term could be found by examining how our main equations (\ref{19}), (\ref{25}), (\ref{26}) and (\ref{27}) would be modified. First, the minimum condition for the new effective potential could simply be found by performing inside (\ref{19}) the following substitution $m^{2}\phi_{0}\rightarrow m^{2}\phi_{0}+I'_{\mathrm{eff}}(\phi_{0},\psi)$, where the prime denotes partial differentiation with respect to $\phi$. Therefore, due to the presence of the huge parameter $\mu^{2}$, as explained in Sec.~\ref{subsec:B}, the solutions $x_{1}$ and $x_{2}$ found there for the variable $x$ would still be valid after performing the previous substitution. Similarly, the effective mass (\ref{25}), the effective potential (\ref{26}), and the partial derivative (\ref{27}) will acquire, respectively, the following new expressions:
\begin{multline}
m^{2}_{\mathrm{eff}}=I''_{\mathrm{eff}}(\phi_{0},\psi)+\left[m^{2}\phi_{0}\rightarrow m^{2}\phi_{0}+I'_{\mathrm{eff}}(\phi_{0},\psi)\right],
\\V_{\mathrm{eff}}\Big|_{\phi_{0}}=I_{\mathrm{eff}}(\phi_{0},\psi)+\left[m^{2}\phi_{0}\rightarrow m^{2}\phi_{0}+I'_{\mathrm{eff}}(\phi_{0},\psi)\right],
\\\frac{\partial V_{\mathrm{eff}}}{\partial R}\Big|_{\phi_{0}}=\left[m^{2}\phi_{0}\rightarrow m^{2}\phi_{0}+I'_{\mathrm{eff}}(\phi_{0},\psi)\right],\nonumber
\end{multline}
where the square brackets mean: the same expressions as those found before but rewritten with the indicated substitution. Now, since, as we saw, the solutions $x_{1}$ and $x_{2}$ one obtains for the variable $x$ from the new minimum condition will remain the same, the effect the radiative corrections will induce on the effective mass will also remain the same as those discussed in Sec.~\ref{subsec:B}. Indeed, as explained below Eq.~(\ref{25}), the latter formula is not sensitive to the initial mass $m$ of the scalar field one starts with, and hence, will also not be sensitive to the substitution $m^{2}\phi_{0}\rightarrow m^{2}\phi_{0}+I'_{\mathrm{eff}}(\phi_{0},\psi)$. Next, the above new expressions for $V_{\mathrm{eff}}\big|_{\phi_{0}}$ and $\partial V_{\mathrm{eff}}/\partial R\big|_{\phi_{0}}$ are the ones that will determine via (\ref{23}) the dynamical equation for the Hubble parameter in the presence of $\phi$-field couplings with matter. However, since the substitution $m^{2}\phi_{0}\rightarrow m^{2}\phi_{0}+I'_{\mathrm{eff}}(\phi_{0},\psi)$ inside (\ref{26}) and (\ref{27}) will not effect the terms containing $\mu^{2}$ (the second and the last two terms in the case of (\ref{26}) and the first term in the case of (\ref{27})), and since only such terms dominate, using the above new expressions for $V_{\mathrm{eff}}\big|_{\phi_{0}}$ and $\partial V_{\mathrm{eff}}/\partial R\big|_{\phi_{0}}$ will not effect our physical conclusions drawn from the dynamical equation (\ref{24}) of the Hubble parameter in the absence of matter couplings with the scalar field $\phi$.

Finally, the fact that not only the mass acquired by the scalar field due to its non-minimal coupling with gravity is huge but the acquired value itself depends on the environment, does not exclude the scalar field from being among the serious candidates for the dark matter sector.
\begin{acknowledgments}
I am grateful to Patrick Labelle for the helpful comments. I would also like to thank the anonymous referee for his/her helpful comments.
\end{acknowledgments}


\end{document}